\documentclass{article} \oddsidemargin=10mm \textwidth=150mm \textheight=240mm \topmargin=-15mm  \newcommand\shortcite\citeyearpar \usepackage{natbib} \renewcommand\cite\citep

\usepackage{epsfig}
\begin{document}

\title
{Earthquake behaviour and large-event predictability in a sheared granular stick-slip system}
\author
{Fergal Dalton$^{1,2}$ and David Corcoran$^1$\\
{\small $^1$Department of Physics, University of Limerick, Ireland; email: david.corcoran@ul.ie}\\
{\small $^2$Dipartimento di Fisica, Universita di Roma "La Sapienza", Italy; email: fergal@idac.rm.cnr.it}}
\date{\today}

\maketitle

\begin{abstract}
We present results from a physical experiment which demonstrates
that a sheared granular medium behaves in a manner analogous to
earthquake activity. The device consists of an annular plate
rotating over a granular medium in a stick-slip fashion. Previous
observations by us include a bounded critical state with a power
law distribution of event energy consistent with the
Gutenberg-Richter law, here we also reveal stair-case seismicity,
clustering, foreshocks, aftershocks and seismic quiescence.
Subcritical and supercritical regimes have also been observed by
us depending on the system configuration. We investigate the
predictability of large events. Using the quiescence between
`shock' events as an alarm condition, it is found that large
events are respectively unpredictable, marginally predictable and
highly predictable in the subcritical, critical and supercritical
states.
\end{abstract}


\section{Introduction}
Due to the destructive nature, and too often devastating
consequences of earthquakes, it has long been the aim of
seismologists to predict them. In this regard, the term
"earthquake prediction" is identified with a short-term process
which would specify the time, place of occurrence and size of a
single earthquake with sufficient accuracy for authorities to
arrange an evacuation~\cite{main-nature}. Yet to date no
undisputed earthquake prediction scheme of this type exists.
Indeed, while the optimism for such a scheme was high in the
1970s~\cite{Geller}, it is  now questionable whether it is
possible at all~\cite{main-nature,Geller,evans}. On the other
hand, "Seismic Hazard Assessment" refers to the process by which
the general likelihood of an earthquake occurring in a region is
established, based on its activity history~\cite{main-1996}.  Such
assessments are generally used for the coding of building
practices for example and may be the best approach for reducing
earthquake damage~\cite{main-nature}.

The predictability issue of earthquakes is intertwined with the
idea that earthquakes are a self-organised critical
phenomenon~\cite{Bak and Tang}. Short-term earthquake prediction
would then be practically impossible, since the sensitivity to
fluctuations at a critical point would require the details of the
entire system to be known. Nevertheless, individual earthquakes
have been associated with precursors, for example foreshocks, or
seismic quiescence preceding an
event~\cite{main-1996,turcotte-1991}, and such phenomena might be
used to signal the imminence of an event. Although, there is no
unambiguous earthquake precursor ~\cite{main-nature,evans},
foreshocks are recognized by the International Association of
Seismology and Physics of the Earth's Interior (IASPEI) as
significant possible precursors to earthquake activity, while
seismic quiescence can currently be neither accepted nor rejected
~\cite{iaspei}.

An important difficulty in the study of earthquakes is that
seismic catalogues on which earthquake statistics and analysis are
based are limited both temporally and spatially. In many parts of
the world, earthquake records are only decades old, while in some
instances increasing and decreasing regional earthquake activity
can result from the installation and removal of monitoring
stations or be a consequence of isolated studies ~\cite{munoz}.
The necessity for synthetic catalogues is clear.

Numerical models capable of generating seismicity catalogues
include notably  the Burridge-Knopoff model and its many
derivatives ~\cite{turcotte-book}. An alternative to the numerical
models, and one which is considered here, is to use a physical
model. Earthquakes on a preexisting fault are assumed  to be
fundamentally the result of many {\em interacting} degrees of
freedom subjected to a slowly increasing shear stress causing
stick-slip motion. Such a definition would also have the necessary
if not perhaps sufficient ingredients for a self-organised
critical process. In this picture higher order features, such as
quiescence, foreshocks, aftershocks and other patterns of
associated earthquake activity would be expected to {\it emerge}
from the system interactions.  The benefit of a physical over a
computational model is that it may more adequately represent or
capture the entire set of physical interaction rules underlying
the system. The distinction between the physical system and the
well-known cellular automaton earthquake model by Olami, Feder \&
Christensen (OFC)~\cite{Olami92}, for example, is then primarily
the rules of interaction.  Generating the seismic activity from
the physical experiment in this non-confined manner, one may
construct synthetic seismicity catalogues and explore these for
evidence of predictability.

The paper is organised as follows. Initially, we outline a
physical experiment already developed by us to study granular
stick-slip motion, and explain why this might be considered as a
model for earthquake activity. We highlight the previous
identification by us of 3 earthquake states, critical, subcritical
and supercritical and then explore the predictability of these
states.

\section{An Earthquake Model}

The possibility that earthquake activity might be related to
granular dynamics has been raised in a number of recent
papers~\cite{nasumo,aharanov,timonen,bean}. We have previously
studied the stick-slip motion of a sheared granular bed
~\cite{dalton-01,dalton-01b} using a physical experiment.  The
experimental apparatus consists of an annular top plate which is
driven over the surface of a granular bed confined to a circular
channel. The plate is driven by the action of a motor via a
torsion spring. In this manner, the motor winds the torsion
spring, increasing the torque on the plate. Ultimately friction
can no longer sustain the applied torque, and the plate spins.

The experiment falls under the general definition for the
earthquake model given above, in which the interacting degrees of
freedom are the positions and velocities of individual grains
determined by the forces, including friction, acting on the
grains. The slowly increasing shear stress is provided by the
action of the motor and torsion spring while the granular medium
is in the solid phase. We consider now the dynamic behavior of the
experiment to see if it is consistent with seismic activity.

We have previously found that the system behaves in a manner
consistent with a self-organized critical process ~\cite{btw} for
a subset of its operating conditions, and we have used the term
"bounded self-organized criticality" to account for the finite
basin of attraction of the critical state~\cite{dalton-01b}. Most
importantly, the distribution of event sizes in the system
critical state is consistent with the Gutenberg-Richter
distribution of earthquakes ~\cite{dalton-01}. Events release
energy E with power-law probability $p(E)\sim E^{-B-1}$, with
$B=0.88\pm 0.04$.  This lies within the
range expected for real earthquakes $1/3 < B < 1$~\cite{Main3}.

The experiment also exhibits subcritical and supercritical states
of earthquake activity\linebreak\cite{dalton-01b}. These states in addition
to the critical state are observed in computational earthquake
models~\cite{RundleKlein}. Indeed, Lomnitz-Adler~\shortcite{Lomnitz}
examined 40 different cases of earthquake cellular automata, and
in all cases the results can be classified into one of these 3
states~\cite{main-1996}.

\begin{figure}
\epsfig{file=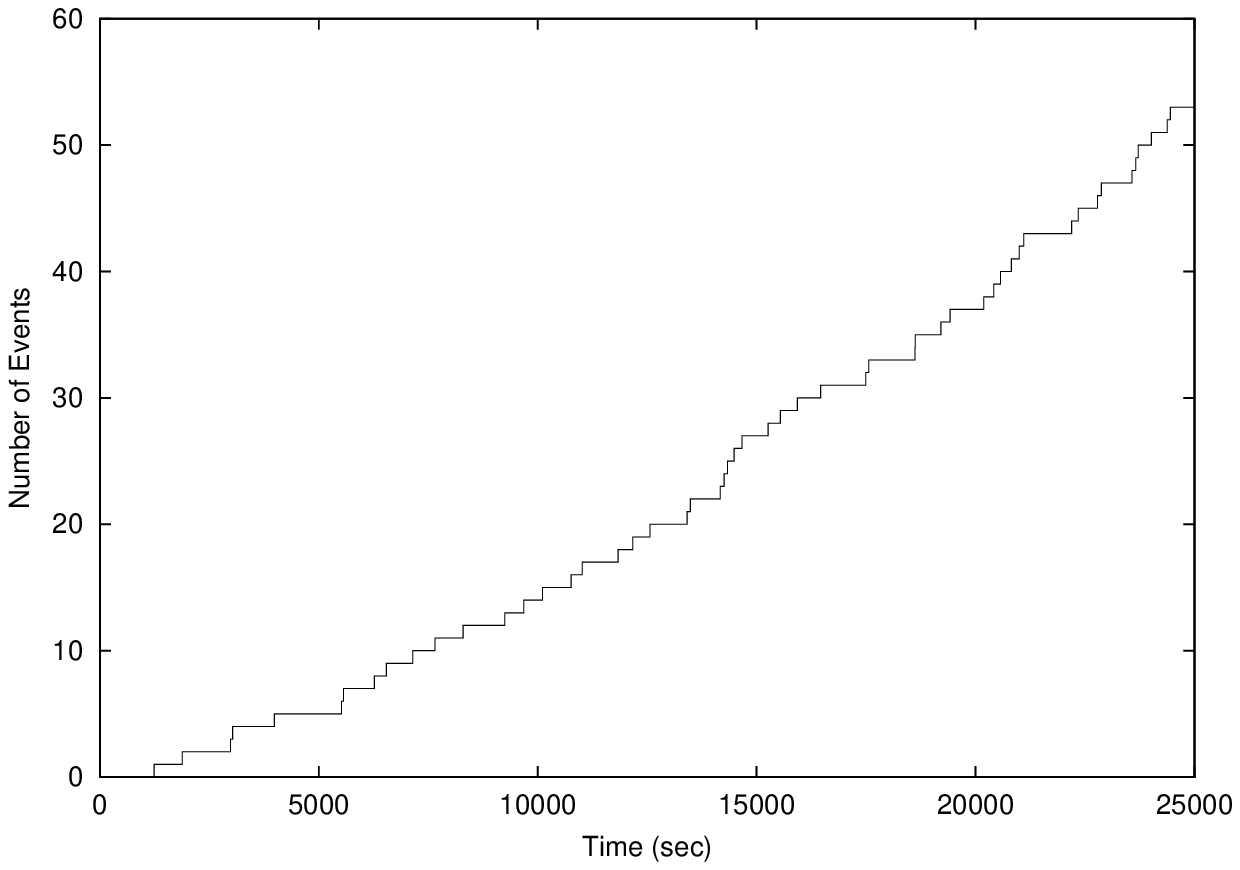,width=85mm}
\centering \caption{\label{fig:omori:cum-quakes} Cumulative number
of events obtained from the apparatus for a critical state. This
data is similar to plots obtained from real earthquake data.}
\end{figure}

\begin{figure}
(a)\epsfig{file=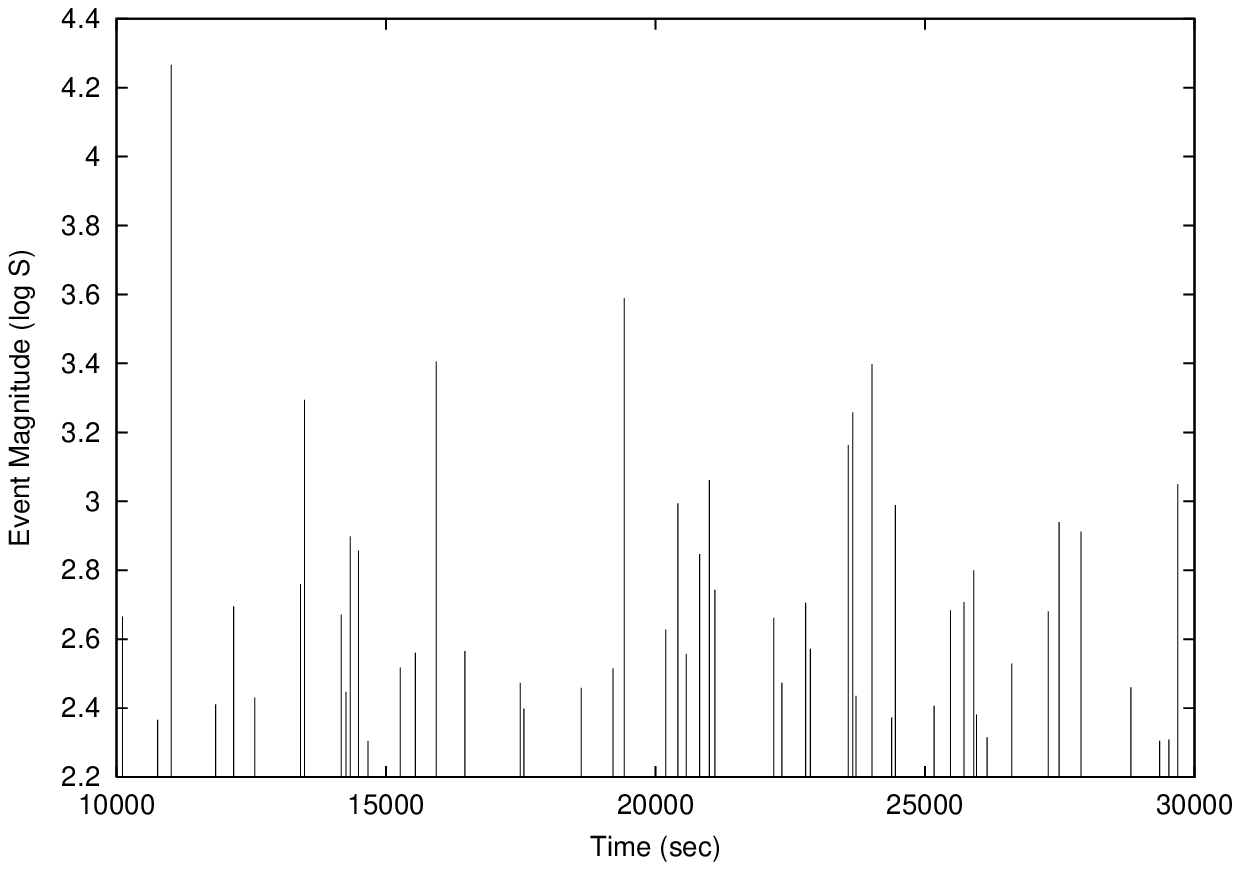,width=85mm}
(b)\epsfig{file=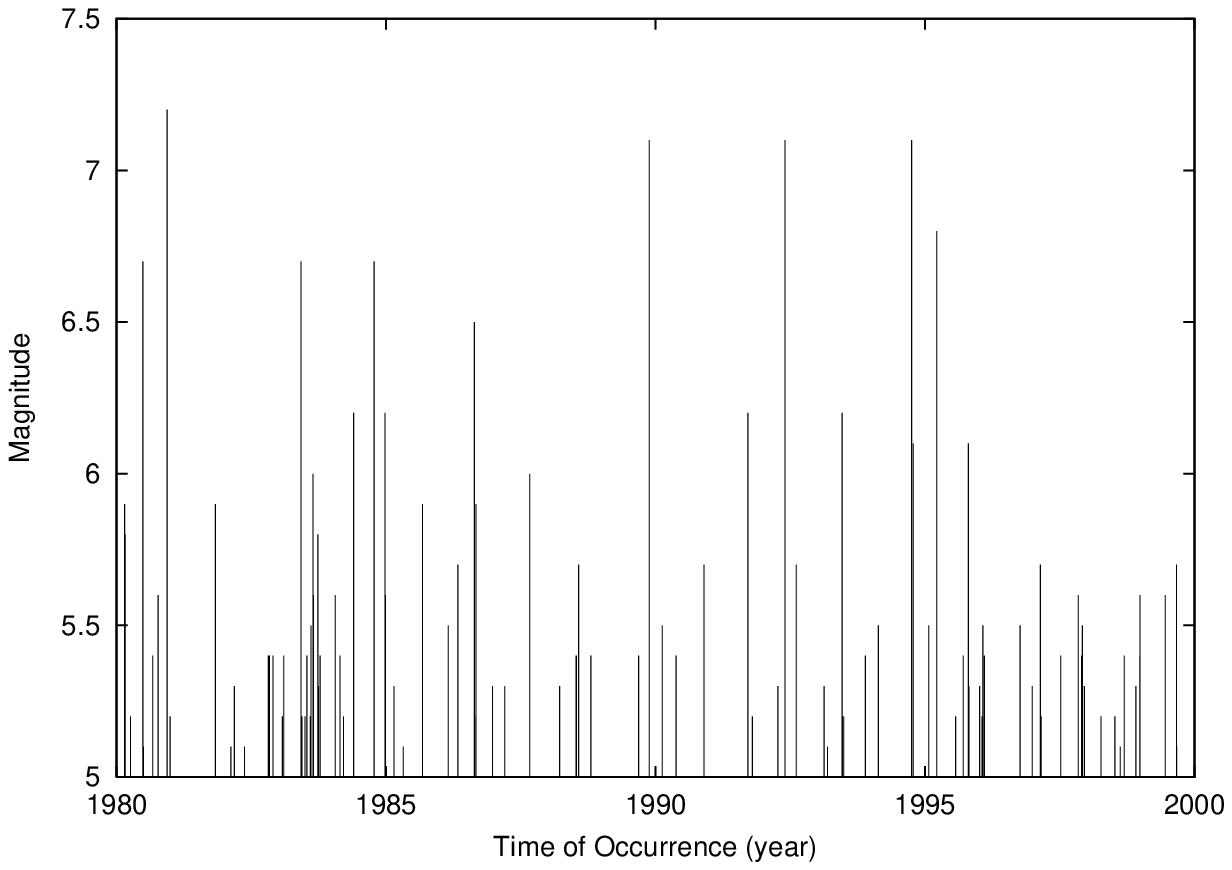,width=85mm}
\centering \caption{\label{fig:omori:big-quakes} (a) the sequence
of large events ($S \geq 10^\circ$) in the critical experiment.
Note the clustering of events. (b) the sequence of large
earthquakes ($M\geq 5.0$) within 500km of San Francisco in a
twenty-year period. Clustering is again apparent. }
\end{figure}

Earthquake activity is also demonstrated in staircase seismicity
and clustering of large events. Large events are defined here for
the physical experiment as those above a threshold size where
deviation from scale-invariant behavior in the power law
distribution of event size occurs (see ~\cite{dalton-01b} ibid.
fig. 2).  Considering a representative critical experiment,
fig.~\ref{fig:omori:cum-quakes} plots the cumulative number of
large events (size $\geq 10^\circ $) in the system as time
progresses. The `staircase' nature of earthquake activity is
clearly shown, and compares well with data of real earthquake
activity in the Mexican Pacific coast~\cite{munoz}. Similar
results are obtained for the supercritical and subcritical states.
The magnitude of all large events in the critical system is
plotted as a function of time in
fig.~\ref{fig:omori:big-quakes}. Activity clearly occurs in
clusters, with isolated events between. Also shown is a plot of
real earthquake activity obtained from the United States
Geological Survey (USGS) web-site~\cite{usgs-search}, for
earthquakes of magnitude $M\geq 5 .0$ in a circle of radius 500~km
about San Francisco. Activity is clustered here also.

The activity of the apparatus, while in the representative
critical state has also been considered in the temporal vicinity
of a large event.  To perform the analysis, the time of occurrence
of all large events was extracted from the data. Activity before
and after each large event was then superimposed, and the average
activity about this 'stacked' or `conglomerate' event computed.
The activity considered was restricted to shock events of size
$S\geq 0.1^\circ $ and was averaged over intervals of 5~s duration. The
time of occurrence of the main event is denoted $t_0$ (in our graphs we set
$t_0=0$). Overall,
there were 104 large events, and the average shock activity was
approximately $R=0.14$ Hz (ie.\ one shock occurred, on average,
every 7 s).

\begin{figure}
\epsfig{file=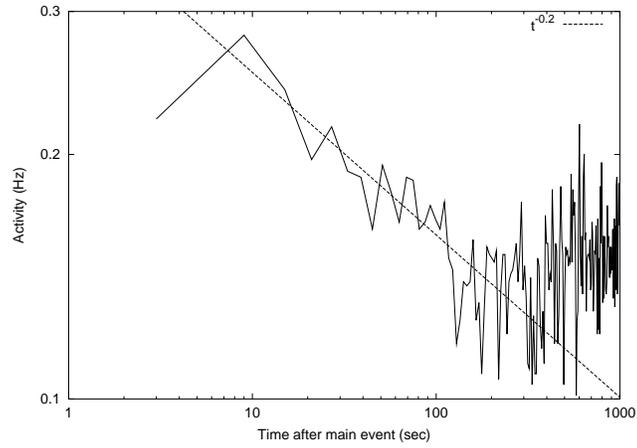,width=85mm}
\centering \caption{\label{fig:omori:10-0.1deg-after} The
aftershock activity averaged for the 104 large events of the
critical experiment. There is a clear elevation in activity
immediately after $t_0$ which decays back to the steady value
$R\simeq 0.14$ Hz.}
\end{figure}

The sequence of activity following the large events is shown in
fig.~\ref{fig:omori:10-0.1deg-after}. Activity is seen to
decrease from an elevated value ($R\simeq 0.28$ Hz), back to the
steady value in a period of 100 to 200 s. The best-fit curve to
this decay is a power-law $R(t) \sim (t-t_0)^{-p}$, shown by the
straight line fit, with exponent $p\simeq 0.2$.  The decay is
reminiscent of Omori's law where the activity of earthquakes falls
off as a power-law with exponent $p \simeq 1$.  Variations in the
power laws exponents for after shock rates have also been obtained
in the computational earthquake model of Hainzl et al.
~\shortcite{hainzl-1999}, where the exponent is dependent on the feed
back of energy lost during an earthquake into the fault zone.

Prior to a large event, the activity decreases from the steady
rate almost to zero, and then immediately before the large event,
increases. Fig.~\ref{fig:omori:10-0.1deg-fore} shows this data.
Activity is largely constant at 0.14~Hz until approximately $t =
-50$~s.  At this point, there is a decrease to 1.9~mHz, and then
an abrupt rise immediately before the main event in a power-law
fashion: $R(t) \sim (t_0-t)^{-q}$, $q\simeq 1.6$. Thus by
examining the sequence of activity about large events in the
physical model, as in the case of earthquakes, aftershocks,
quiescence, and foreshocks are also observed.

In the physical experiment of a sheared granular medium one
therefore sees many of the characteristics of earthquake dynamics,
including a power law distribution of event energy consistent with
the Gutenberg-Richter law, stair-case seismicity, clustering,
foreshocks, aftershocks and seismic quiescence. We conclude that
the physical experiment is behaving in a manner analogous to an
earthquake fault.

\section{Predictability of the Earthquake Model}

\begin{figure}
\epsfig{file=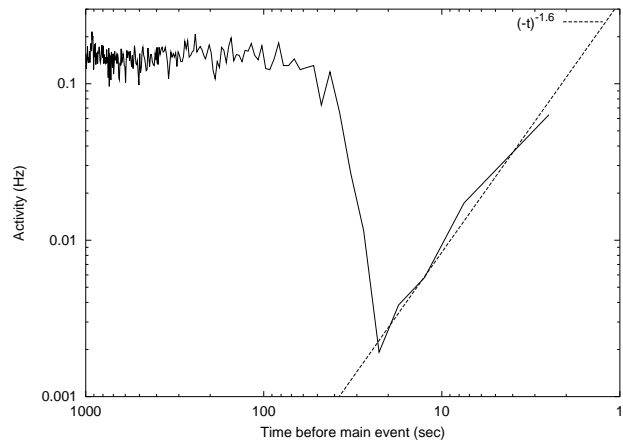,width=85mm}
\centering \caption{\label{fig:omori:10-0.1deg-fore} The averaged
precursory activity before the 104 large events of the critical
experiment. Activity decreases from the steady value and rapidly
rises before the main event.}
\end{figure}

Given the presence of precursors, what are the implications for
predictability in the representative critical state? The power-law
increase in activity of fig.~\ref{fig:omori:10-0.1deg-fore} is
based on 104 large events for which 48 foreshocks were issued.
However, instead of 48 of the 104 large events issuing one
foreshock each, the foreshocks are due only to 25 of the large
events (24\%).  Despite the clear change in statistical behavior
as a large event is approached, the mean behavior observed
describes only a small proportion of large events and thus
severely limits its application for prediction purposes. It is
interesting that studies of real earthquakes indicate that
approximately one-quarter have
foreshocks~\cite{turcotte-1991,turcotte-book}, in agreement with
the value of 24\% obtained here. Comparison of greater volumes of
real and artificial data are needed to establish if this agreement
holds, or is merely coincidence.

Consider now the quiescence of
fig.~\ref{fig:omori:10-0.1deg-fore} where the activity before
the main shocks decreases to 1.9~mHz. This indicates that, for the
5~s interval represented by that point ($-25\leq t_0-t \leq -20$
s), only one of the 104 large events was active.  This is clearly
beyond the bounds of random probability, and hence it must be
concluded that large events are almost always preceded by a quiet
period 20 to 25~s before a large event. Unfortunately, the sense
of this conclusion cannot be reversed. That is, it is not correct
to say that a quiet 5~s interval will lead to a large event in
about 20~s time.  The mean rate of shocks is 0.14 Hz,
so two in every seven 5~s intervals will, on
average, be inactive. Hence a more detailed study of each sequence
of activity is necessary.

Analysis of the data reveals that 84 of the 104 events were
shortly preceded by a 40~s period in which no activity occurred
and for the entire duration of the experiment, there were a total
of 160 periods lasting 40~s each, in which no activity took place.
Again, 84 of these led to a large event.   We can use this to
generate a simple prediction algorithm where a 40~s interval in
which no activity occurs is the condition upon which an "Imminent
Large Event" alarm is raised. We stress that we are not proposing
a novel earthquake prediction scheme here, merely presenting an
analysis of our data. The 20 large events which were not preceded
by an alarm can be regarded as a failure-to-predict. Furthermore,
there were 76 false-alarms, also failures of the scheme, and there
are a total of 84 successful predictions.  Hence the success rate
of this scheme is 84/(76+84+20) = 47\%.  The false alarm rate, the
fraction of alarms which are not followed by a large event, is
76/160 = 47\% and the failure-to-predict rate, the fraction of
large events which are not predicted, is 20/104 = 19\%.

\begin{figure}
\epsfig{file=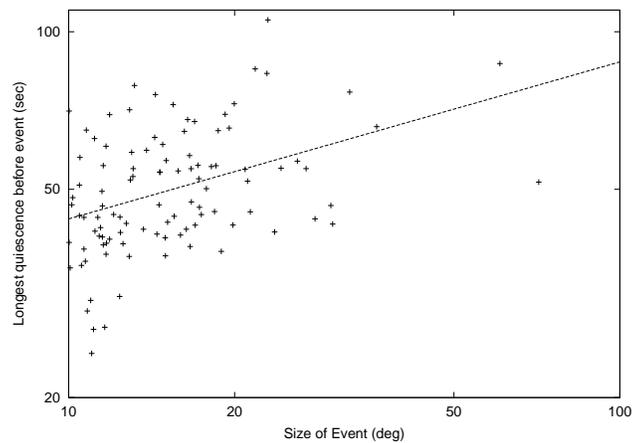,width=85mm}
\centering\caption{\label{fig:omori:intervalVsize}
The longest interval preceding a large event plotted against the
size of the event.}
\end{figure}

Finally, the relationship between the duration of the quiescent
period before an event to the subsequent size of the main event,
is presented in fig.~\ref{fig:omori:intervalVsize}. The
power-law bestfit indicates an upward trend, though the data is
far too poorly correlated to make any definite assumptions
(correlation coefficient $r=0.40$). Similar results are observed
for earthquakes~\cite{schol}. The implication is that event size
is effectively independent of the quiescence duration. While it
might be possible to predict the imminent arrival of an event, the
conclusion that must be drawn is that it is not possible to
predict its size. One notes this result differs from recent
computational work~\cite{Hainzl2}, in which quiescence in a
modified OFC cellular automata model was shown to correlate with
subsequent event size.

\section{Other Earthquake States}

\subsection{Subcritical State}

\begin{figure}
\epsfig{file=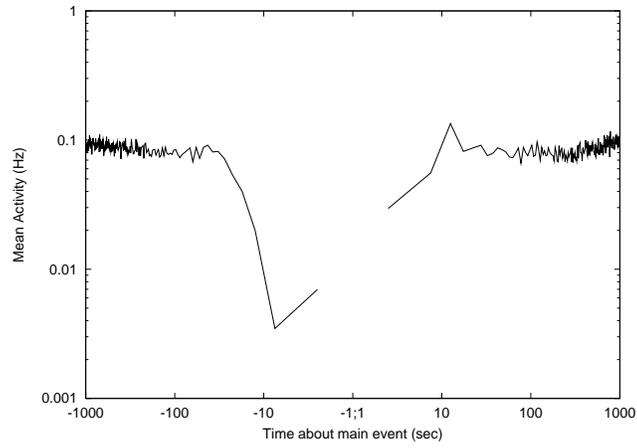,width=85mm}
\centering \caption{\label{fig:omori:hovered-foreshocks} The
activity about large events in a subcritical state.  A precursory
quiescence is evident, but no foreshocks or aftershocks.}
\end{figure}

Fig.~\ref{fig:omori:hovered-foreshocks} is a plot of activity
about `large' events (defined here as event sizes $\geq 3^\circ $)
for a representative subcritical experiment. While the device
still demonstrates seismic quiescence, foreshocks are not so
evident and a decaying aftershock sequence is not present.
Overall, in the subcritical state, it emerges that large event
predictability based on quiescence is low. The optimum algorithm
using a 20~s interval predicts large events with a 23\% success
rate, 64\% false-alarms and 61\% failure-to-predict.

\subsection{Supercritical State}

The seismicity of large events in a representative supercritical
state is dominated by the recurrent large events to which this
state is susceptible~\cite{dalton-01b}. The events do not appear
to cluster to one another.  For the supercritical state, it may be
true to say that the next large event ($S\geq 15^\circ $) will
probably occur 200~s after the previous one. Interestingly, the
torque fluctuations for supercritical data do not reveal discrete
Fourier components in their power spectra~\cite{dalton-01b},
though this may arise from the periodic element being lost in the
dominant 1/f$^2$ noise spectrum.

\begin{figure}
\epsfig{file=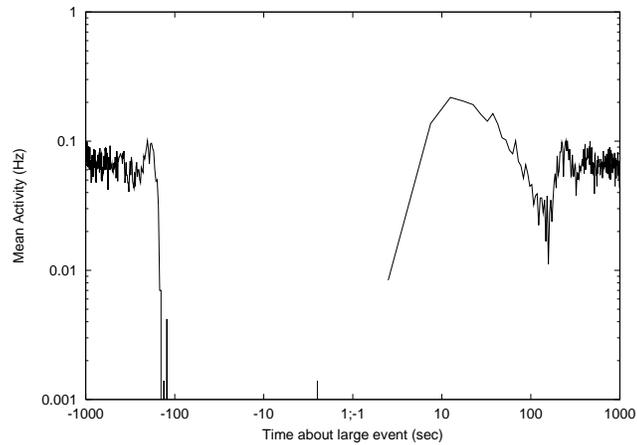,width=85mm}
\centering \caption{\label{embedded-foreshocks} The averaged
sequence of shocks before and after large events in the
supercritical state.  Activity beforehand drops from approximately
65 mHz to zero at $t=-150$~s, with only a single foreshock
occurring at t= -2.5 s.}
\end{figure}

Activity about the large events in the supercritical state is
notably different to the previous two states (see fig.
~\ref{embedded-foreshocks}).  A precursory quiescence is still
observed, though in this case, the quiescence commences at
$t=-150$~s, from 0.07 Hz to zero.  The zero activity is completely
maintained right up to the main event excluding one shock at $t =
-2.5$~s. Thus foreshocks appear to be non-existent in this state.

Aftershocks also behave in a different fashion to the critical and
subcritical states. The rate of shocks is seen to decrease after
the main event, following an exponential decay, to a level {\it
below} the mean rate, and then, with decaying oscillations, it
settles back to the mean rate.  The existence of the exponential
decay suggests that the apparatus is no longer operating at a
critical point characterized by scale invariant power-laws. The
oscillatory effect is attributed to the recurrent nature of these
large events.

In the supercritical state, large events are almost perfectly
predictable as the following data will demonstrate.  Over the
period of the experiment there were 142 large events and 144
intervals of length 100~s or longer.  Each of these large events
was preceded by such an interval, and so a prediction based on
these parameters will have a 99\% success rate, 1\% false alarms
and 0\% failures-to-predict.

\section{The General Critical State}

The critical behavior marks a transition between the well defined
sub and supercritical states. The latter are easily identifiable
and experimentally repeatable, but the critical state has been
found to exhibit a richer variety of behavior~\cite{dalton-01b}. Of
seven critical experiments, all exhibit qualitatively similar
patterns in terms of the clustering and intermittency of large
events. However, some of them have fewer large events (ie.\ $S\geq
10^\circ $) and so it is necessary to reduce the threshold for
`large' events in order to make the clustering visible.  The
cumulative number of events obtained from four of the seven
experiments tend to show a more uniform slope, attributed to this
rarity of large events; rarer large events necessarily implies
more frequent smaller events, leading to the more uniform slope.

The Omori-type power-law decay in aftershocks is repeated by five
of the seven experiments, with exponent $0.2\leq p \leq 0.45$. For
precursory behavior, all show a quiescence beginning approximately
40~s before the main event, followed by an increase in activity in
six experiments.  However, the increase can only be clearly fit
with a power-law in three cases, the remainder having statistics
that are too low for any curvefit (with perhaps only 10 foreshocks
occurring in total).

Of the seven experiments, three had similar success rate
$\simeq$47\%, false alarm rate $\simeq$48\% and failure-to-predict
rate $\simeq$19\%.  The remaining four had success rate
$\simeq$16\%, false alarm rate $\simeq$80\% and failure-to-predict
rate $\simeq$44\%.  This disparity may arise because the
experiments with higher predicability are nearer the transition to
the supercritical state where events are highly predictable. The
experiments with lower predictability may similarly be near the
transition to the subcritical state where large events are highly
unpredictable. We note that this interpretation would also be
consistent with variability in the torque value, and variability
in the event size and duration distributions previously reported
for the critical state~\cite{dalton-01b}.

\section{Conclusion}
A sheared granular experiment has been shown here to behave in a
manner analogous to earthquake activity and has previously been
shown by us to exhibit bounded self-organised criticality. The
observation of 3 distinct dynamic states, above, below and in a
critical region has an important implication for the concept of
self-organized criticality as applied to physical systems,
specifically here earthquakes. The sub and supercritical states
are outside the critical region, yet in these cases and at
criticality, for many different initial experimental
configurations scale invariant event distributions are observed
over several decades. Thus, while not perhaps self-organised
critical, the experimental system is robustly self-organised to
near criticality. It may be that earthquakes are similarly
self-organised.

In the physical experiment, the ability of the granular material
to dilate is believed to be responsible for the tuning of the
system to the different system states~\cite{dalton-01,dalton-01b}.
A physical cause is therefore responsible for the observation of
the different states. In actual earthquake statistics,  the
subjective choice of study area is known to effect the tail of the
distribution, with large areas revealing sub-critical behavior and
smaller areas more supercritical behavior ~\cite{main-1996}. This
points to the difficulty of objectively constraining the
observational data for the largest earthquakes. By studying the
limits and sizes of the basins of attraction for the various states of the
physical experiment, one might be able to determine the most
probable tail for the scale-invariant distribution.  Preliminary
results ~\cite{LynchCD} suggest that the most common state for the
experiment is subcritical, in agreement with recent statistical
work on global earthquake data~\cite{Leonard}.

Seismic quiescence is common to both earthquakes and large events
in the experiments here. This quiescence has been utilized as a
simple prediction scheme. The prediction algorithm essentially
obtains the {\em statistical} likelihood of a large event
occurring after a long quiescent period. From this {\em
statistical} point of view, a certain level of prediction will
always be possible even though the system is close to criticality,
assuming a detailed history of the fault is available and the
behavior is stationary.

The simple prediction algorithm implemented suggests marginal
predictability for the critical state, with some experiments
predictable and some not, the difference being the proximity of
the specific critical state to the sub and super critical regimes.
Large events in the subcritical state seem to exhibit low
predictability, while events in the supercritical state appear to
be (almost) perfectly predictable.

This research has been funded by the University of Limerick
Foundation and Enterprise Ireland.  We are grateful to Gerry Daly
for devoting considerable time and effort to this project, and
acknowledge Ian Clancy and Christina Conway for useful
discussions. Finally we would like to thank Ian Main for his
helpful comments and suggestions while writing this paper.


\newpage

\end{document}